\documentclass[12pt]{article}

\usepackage{epsfig}
\usepackage{amsmath}
\usepackage{amssymb}
\usepackage{graphicx}
\usepackage{cite}
\def\independenT#1#2{\mathrel{\setbox0\hbox{$#1#2$}%
 \copy0\kern-\wd0\mkern4mu\box0}} 

\def\be{\begin{equation}}
\def\ee{\end{equation}}
\def\bea{\begin{eqnarray}}
\def\eea{\end{eqnarray}}
\def\bsp{\begin{split}}
\def\esp{\end{split}}
\begin{document}
%\titlepage
\titlepage
\begin{flushright}
IFJPAN-IV-20011-3 \\
\end{flushright}
\vspace*{1in}
\begin{center}
{\Large \bf Gluon saturation and entropy production in proton proton collisions.}\\
\vspace*{0.5cm}
 Krzysztof Kutak \\
\vspace*{0.5cm}
 {\it Instytut Fizyki Jadrowej im H. Niewodniczanskiego,\\
Radzikowskiego 152, 31-342 Krakow, Poland\\}
\end{center}
\vspace*{1cm}
\centerline{(\today)}

\vskip1cm
\begin{abstract}
We study properties of high energy factorisable gluon densities focusing on saturation effects.
In particular we show that the property of saturation of unintegrated gluon density allows for introduction of 
thermodynamical entropy associated with production of gluons. This is due to the observation that saturation scale acts like a mass of a gluon which related with temperature via thermodynamical relations allows for calculations of entropy. 
We also show that obtained entropy behaves like multiplicity of produced gluons.
\end{abstract}
%%%%%%%%%%%%%%%%%%%%%%%%%%%%%%%%%%%%%%%%%%%%%%%%%%%%%%%%%%%%%%%%%%%%%%%%%%%% 
\section{Introduction}
Perturbative Quantum Chromodynamics (pQCD) at high energies generates scale called saturation scale $Q_s$.
The need for the existence of such a scale originates back to investigations of unitarity violation \cite{Gribov:1984tu} by linear equations 
of pQCD \cite{Kuraev:1977fs,Balitsky:1978ic}.
To resolve this problem, higher order perturbative corrections of nonlinear type within the Balitsky-Kovchegov, CGC/JIMWLK  framework (i.e. nonlinear modifications to summation of logarithms of energy $\alpha_s^n(\ln s)^n$) were considered \cite{Balitsky:1995ub,Kovchegov:1999yj,Kovchegov:1999ua,JalilianMarian:1997jx,JalilianMarian:1997gr,JalilianMarian:1997dw,JalilianMarian:1998cb,Kovner:2000pt,Weigert:2000gi,Iancu:2000hn,Ferreiro:2001qy}. These unitarity corrections have a clear physical meaning for they allow gluons to recombine what at some point balances gluons splitting. The saturation scale depends on energy and its existence prevents gluon densities from rapid growth. Existing data suggest that the phenomenon of saturation occurs in nature. The seminal example is provided by a discovery of the geometrical scaling in HERA data \cite{Stasto:2000er} and more recently by a presence of geometrical scaling in production of inclusive jets in the LHC data \cite{McLerran:2010ex,Praszalowicz:2011tc}.\\
Here, we would like to use the saturation  property of gluons to derive an expression for amount of thermodynamical entropy associated with generation of saturation scale $Q_s$ which acts as effective mass of gluon during collision of hadrons. We show that such a construction is possible due to the relation of temperature to the saturation scale and the property of saturation scale which acts as an effective mass of gluon. Our main result are formulas (\ref{eq:entropia}), (\ref{eq:finalfinal}), for other approaches to entropy production we refer the reader to \cite{Blaizot:2005wr,Bialas:2006mg,Blaizot:2006tk,Lublinsky:2007mm,Gubser:2008pc}. 
We also show that the number of gluons whose distribution is given by unintegrated gluon density (which in perturbative framework is obtained by summing  via integral equations Feynman diagrams describing emissions of gluons) (\ref{eq:numberofgluons}) is up to a constant factor equal to number of produced gluons therefore it can be directly linked to entropy of produced gluons. This is possible because of universality of saturation scale which is the only scale in the problem and which links those quantities in a unique way.  \\
The letter is organised as follows. In section 2 we obtain a formula for a gluon density in the adjoint representation of the color group and we show that saturation is reached at larger scales than in the fundamental representation. In section 3 we use the obtained gluon density to calculate the cross section for inclusive gluon production and we show that the cross section behaves like the saturation scale. In the 4'th section we derive formula for entropy associated with produced gluons and we relate it to number of gluons given by unintegrated gluon density. Finally we conclude that since saturation scale saturates itself the entropy of gluons coming from saturated region has to be bounded.  

\section{Saturation scale in the adjoint color representation}
\begin{figure}[t!]
  \begin{picture}(30,30)
    \put(30, -80){
      \includegraphics{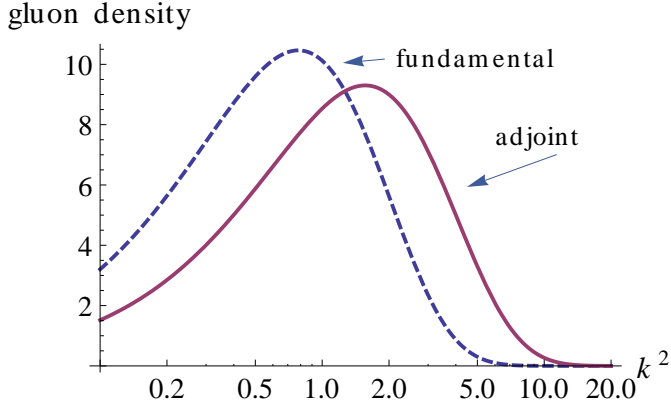}
    }

      \end{picture}
\vspace{3cm}
\caption{\em \small High energy factorisable gluon density. The maxima signalise existence of the saturation scales. Magenta continuous line represents the gluon density in the adjoint representation. The blue dashed line represents the gluon density coming from the dipole in the fundamental representation. Plot is made for $x=10^{-4}$.}
%\vspace{7cm}
\label{fig:plotvel}
\end{figure}

The high energy factorisable \cite{Catani:1990eg} unintegrated gluon density can be linked to a dipole amplitude for a color dipole scattering off a hadronic target via the following relation \cite{Kovchegov:2001sc,JalilianMarian:2005jf,Kharzeev:2003wz}:
\begin{equation}
\phi_{(G),(Q)} (x,k) \, = \, \frac{C_F}{\alpha_s \, (2 \pi)^3} \, \int d^2 \mathbf b \, 
d^2 \mathbf r \, e^{- i \mathbf k \cdot \mathbf r} \ \nabla^2_r \, N_{(G),(Q)} (r,
b, x).
\label{eq:transf}
\end{equation}
where the subscripts $Q, G$ specify whether considered dipole is in the fundamental or in the adjoint color representation. The impact parameter at which a dipole collides with the hadron is  $\mathbf b$, the size of the dipole is $r\!=\!|\mathbf r|$ and $k\!=\!|\mathbf k|$ is its conjugated variable characterising transversal momentum of the gluon. The two dimensional vectors  $b\!=\!|\mathbf b|$, $k\!=\!|\mathbf k|$, $r\!=\!|\mathbf r|$ lay in the transversal plane to the collision axis. The variable $x$ is related to the longitudinal momentum component of the gluon momentum.\\ 
Due to the fact that in an electron-proton (e-p) collisions a dipole originates from a photon dissociation the associated gluon density is taken to be in the color fundamental representation while in hadron-hadron (A-A, p-A) collisions it originates most likely from gluon, therefore the associated gluon density with a gluon dipole is in the color adjoint representation. The relation between these two amplitudes in the large $N_c$ limit is the following \cite{Kovchegov:2001sc} :
\begin{equation}
  N_G ({r}, { b}, y) \, = \, 2 \, N_Q ({ r}, {b}, y) - N_Q^2
  ({ r}, {b}, y),
\label{eq:2NN}
\end{equation}
In our investigations of the gluon density in the adjoint representation, we use the one which can be obtained from the GBW model \cite{GolecBiernat:1998js} for the dipole amplitude. This model although, quite simplistic, is a nice laboratory for studying physics of saturation. In particular, we find that the integrals needed in investigated in the next section inclusive gluon production can be performed exactly.
The GBW amplitude reads:
\be
N_Q(x,r,b)=\theta(b_0-b)\Bigg[1-\exp\left(-\frac{r^2 Q_s^2(x)}{4}\right)\Bigg]
\label{eq:GBW98}
\ee  
where $Q_s(x)$ is the saturation scale and it is modelled to be $Q_s(x)\!=\!Q_0\left(\frac{x_0}{x}\right)^{\lambda/2}$ where $\lambda$, $x_0$,  are free parameters and $b_0$ defines proton's radius. This amplitude saturates for large dipoles $r\!>\!\!>\!\!2/Q_s$ and exhibits geometrical scaling which has been confirmed by data \cite{GolecBiernat:1998js,Stasto:2000er}.
The unintegrated gluon density in the adjoint representation can be obtained in a straightforward way via the transformation (\ref{eq:transf}).
However, before we perform this transformation, let us recall the result for the gluon density in the fundamental representation:
\begin{equation}
\phi_Q (x,k)=\frac{N_c A_{\perp}}{2\pi^2\alpha_s}\frac{k^2}{Q_s(x)}\exp\left[-\frac{k^2}{Q_s^2(x)}\right],
\label{eq:ugd2}
\end{equation}
where we assumed a cylindrical shape of the proton which integrated over the impact parameter $b$ gives the transversal area $A_{\perp}=\pi R^2$, where $R$ is a radius of a proton.
Now motivated by the fact that this function exhibits a maximum as a function of $k^2$, see fig. (\ref{fig:plotvel})
for fixed $x$, we define the saturation scale as momentum for which the $\phi_Q(x,k)$ has maximum as a function of $k^2$:
\be
Q_s(x)\equiv\partial_{k^2} \phi_Q(x,k)=0
\label{eq:poch2}
\ee
Applying this condition we obtain:
\be
\partial_{k^2}\phi_Q(x,k)=\frac{\phi_Q(x,k_t)(Q_s^2(x)-k_t^2)}{k^2Q_s^2(x)}=0
\label{eq:poch2}
\ee
This observation has been used in \cite{Kutak:2009zk} to formulate the GBW model as a solution to a transport equation of a form: 
\begin{equation}
\partial_Y \phi(Y,L)+\lambda\partial_L \phi(Y,L,b)=0
\label{eq:trans}
\end{equation}
with initial condition $\phi(k^2,x_0)=k^2 \exp(-k^2)$
where $Y=\ln 1/x$ and $L=\ln k^2$.
Now we calculate the saturation scale for gluons in the color adjoint representation. To see that the effective saturation scale for gluons in the adjoint representation is larger than in the fundamental representation we write the generic expression for the dipole cross section which saturates at large dipole size as:
\be
N_Q(x,b,r)=1-e^{-f(x,r,b)}
\label{eq_gludipol}
\ee
Using (\ref{eq:GBW98}) in (\ref{eq:2NN}) and rearranging terms we obtain:
\be
N_G(x,b,r)=1-e^{-2f(x,r,b)}
\ee
where the factor $2$ in the exponent is the origin of the difference between the saturation scale for gluon density in the adjoint representation and gluon density in the fundamental representation.
To obtain the gluon density we use (\ref{eq:transf}) together with (\ref{eq:2NN}) to obtain gluon density which reads:
\be
\phi_G(x,k)=\frac{C_FA_{\perp}}{4\pi^2\alpha_s}\frac{k^2}{Q_s^2}\exp\left[-\frac{k^2}{2Q_s^2(x)}\right]
\label{eq:adjointglue}
\ee
Applying our definition for saturation scale and calculating derivative we find:
\be
\partial_{k^2}\phi_Q(x,k^2)=\frac{\phi_G(x,k)(2Q_s^2(x)-k^2)}{2k^2 Q_s^2(x)}
\label{eq:poch2}
\ee
what means that the effective saturation scale for gluons in the adjoint representation is larger than in the fundamental representation
and reads $(Q_s^2)_G=2(Q_s^2)_Q$.
%%%%%%%%%%%%%%%%%%%%%%%%%%%%%%%%%%%%%%%%%%%%%%%%%%%%%%%%%%%%%%%%%%%%
\section{Inclusive gluon production}
In this section we calculate the rapidity dependence of the cross section for the inclusive gluon production using the gluon density (\ref{eq:adjointglue}). Since we will use only the gluon density in the adjoint representation we skip the index $G$. Here, we consider a slightly asymmetric configuration near the central region, that is why we use the same parameterisation for the gluon densities but with two different saturation scales. In the following, we are interested in qualitative results, therefore for simplicity, we assume hadrons to be protons. Following \cite{Kharzeev:2004if}, we use  $x_1=(Q_{s\,1}/\sqrt{s})e^{y}$ and $x_2=(Q_{s\,2}/\sqrt{s})e^{-y}$ where $\sqrt{s}$ is the energy of the collision.
The formula for inclusive gluon production reads \cite{Gribov:1984tu,Kovchegov:2001sc,Braun:2000bh,Kharzeev:2003wz,Kovner:2006wr}:
\begin{equation}
\frac{d \sigma}{d^2 p_t \, dy} \, = \, \frac{2 \, \alpha_s}{C_F} \,
  \frac{1}{{p_t}^2} \, \int d^2 k \, \phi({ k}, x_1) \, \phi
  ({{k} - p_t}, x_2)
\label{eq:inclusiveg}
\end{equation}
For recent theoretical studies of the inclusive gluon production with the NLO effects coming from the running coupling constant we refer the reader to \cite{Horowitz:2010yg}. For recent phenomenological works see \cite{Levin:2010br,Levin:2010zy,ALbacete:2010ad,Albacete:2010bs,Tribedy:2011yn,Tribedy:2010ab,Dumitru:2010iy,Dusling:2009ni}.

Using the expressions for the unintegrated gluon densities and performing the integral over $d^2k$, we obtain:
$$
\frac{d\sigma}{dy\,d^2p_t}=\frac{A_{\perp}^2C_F Q_{s\,1}^2 Q_{s\,2}^2 e^{-\frac{p_t^2}{Q_{s\,1}^2+Q_{s\,2}^2}}}{2 \pi ^3\alpha_s %p_t^2
   (Q_{s\,1}^2+Q_{s\,2}^2)^5}
(p_t^4 Q_{s\,1}^2 Q_{s\,2}^2+p_t^2 (Q_{s\,1}^2-Q_{s\,2}^2)^2 (Q_{s\,1}^2+Q_{s\,2}^2)%\cite{Kovchegov:1999yj}
$$
\be
+2 Q_{s\,1}^2 Q_{s\,2}^2
   (Q_{s\,1}^2+Q_{s\,2}^2)^2))
\label{eq:rapdependence}
\ee
Integrating over $d^2p_t$ and introducing minimal $p_{t\,min}$ because the integral is logarithmically divergent, we obtain:
\be
\frac{d\sigma}{dy}=\frac{2A_{\perp}^2 C_F Q_{s\,1}^2 Q_{s\,2}^2 e^{-\frac{p_{t\,min}^2}{Q_{s\,1}^2+Q_{s\,2}^2}} \left(p_{t\,min}^2 Q_{s\,1}^2 Q_{s\,2}^2+Q_{s\,1}^6+Q_{s\,2}^6\right)}{\pi ^2 \alpha_s (Q_{s\,1}^2+Q_{s\,2}^2)^4}
+
 \frac{4A_{\perp}^2 C_F Q_{s\,1}^4 Q_{s\,2}^4 \Gamma \left(0,\frac{p_{t\,min}^2}{(Q_{s\,1}^2+Q_{s\,2}^2)}\right)}{\pi ^2 \alpha_s
   (Q_{s\,1}^2+Q_{s\,2}^2)^3}  
\label{eq:crossrapid}
\ee
Assuming, that the saturation scale in one of the colliding hadrons is the smallest scale in the problem $p_{t\,min}\!\simeq\!Q_{s\,1}\!\!<\!\!Q_{s\,2}$ we perform expansion in terms of $Q_{s\,1}/Q_{s\,2}$ in (\ref{eq:crossrapid}) arriving at:
\begin{equation}
\frac{d\sigma}{dy}=\frac{2 A_{\perp}^2 C_F Q_{s\,1}^2}{\pi ^2 \alpha_s}-
\frac{2 A_{\perp}^2 C_F Q_{s\,1}^4}{\pi ^2 \alpha_s
   Q_{s\,2}^2}\left(2 \log \left(\frac{Q_{s\,1}^2}{Q_{s\,2}^2}\right)+5+2 \gamma_E\right)+O\left((Q_{s\,1}^2/Q_{s\,2}^2)^3\right)
\label{eq:finaleq}
\end{equation}
$$
$$
A comment regarding formula (12) is in order. In derivation of (12) it has been assumed that one of gluon densities is in a dilute regime, the more general formulation which would allow for taking into account interaction of two dense targets would  most probably introduce regulator of the infrared divergence. However, such framework at present is not known and we regulate the divergence by the lowest physical scale in the problem.
The result above shows in agreement with \cite{Horowitz:2010yg,Levin:2010br} that $d\sigma/dy$ behaves like the saturation scale making this observable particularly interesting for studying physics of dense partonic systems formation. On Fig. (\ref{fig:plotvel}) we show results obtained from formulas 
(\ref{eq:crossrapid}) and (\ref{eq:finaleq}). We see that the full result is well approximated by the first term if the saturation scales are significantly different.
%%%%%%%%%%%%%%%%%%%%%%%%%%%%%%%%%%%%%%%%%%%%%%%%%%%%%%%%%%%%%%%%%%%%
\begin{figure}[t!]
  \begin{picture}(30,30)
    \put(30, -80){
      \includegraphics{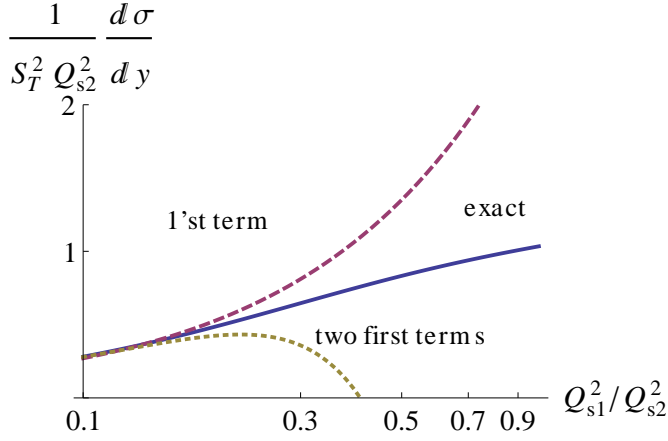}
    }

      \end{picture}
\vspace{3cm}
\caption{\em \small Cross section for inclusive gluon production (normalised to larger of two saturation scales and square of the transversal area $A_\perp\approx S_\perp$) as a function of ratio of saturation scale. The blue line is the full cross section (\ref{eq:crossrapid}), the magenta line is the leading term, the green is the contribution from the sum of leading and subleasing term.}
%\space{7cm}
\label{fig:plotted}
\end{figure}
\section{Entropy}
Now we would like to relate the obtained cross section for the inclusive gluon production to produced entropy in the collision due to emergence of effective mass of gluons during the collision. 
We wish, however, to consider dense system where the following formula is not easily justified :
\be
S=\int\frac{d^3z\,d^3k}{(2\pi)^3}[-f(z,k)\ln f(z,k)+s_I(1+s_if(z,k))\ln(1+s_if(z,k))]
\ee  
where $f(z,p)$ is phase space density of patrons/hadrons, $s_i=-1$ for fermions and $s_i=1$ for bosons. 
We see the possibility to address the question of entropy of dense system, following ideas of \cite{Kharzeev:2005iz,Castorina:2007eb} where relation between saturation scale and temperature of thermally produced hadrons has been obtained within partonic picture of QCD and it reads:
\be
T=\frac{Q_s(x)}{2\pi}
\label{eq:temperatura}
\ee
where $x$ stands for a longitudinal momentum fraction of hadron carried by a parton. 
The formula above is a direct consequence of the Unruh effect \cite{Unruh:1976db} within QCD. The Unruh effect \cite{Unruh:1976db} states that accelerated observer in its rest frame feels thermal radiation of temperature $T=\frac{a}{2\pi}$. In the construction leading to QCD application, a decelerating off shell gluon with distribution in transversal momentum given by unintegrated gluon distribution is considered. Due to deceleration, a horizon is formed tunnelling through which produces a thermal ensemble with a density matrix of the Maxwell-Boltzmann type. To determine the deceleration the authors of \cite{Kharzeev:2005iz} use the Wong equations \cite{Wong:1970fu} for color particle moving in a homogeneous chromoelectric field. Solutions of this equations give a result that $|a|=Q_s$ which leads to
eq. (\ref{eq:temperatura}).     
Furthermore, it is well known that saturation scale effectively acts as a regulator of the infrared divergent behaviour of the unintegrated gluon density. At the lowest order in QCD (in phase space region where the total energy is larger than any other scale) it is:
\be
\phi=\frac{\alpha_s C_F}{\pi}\frac{1}{k^2}
\ee
This quantity has a structure of propagator and we see that at small values of $k$ it becomes divergent.
Higher order perturbative corrections coming from BFKL resumation do not solve this problem. This low $k$ behaviour of BFKL equation is interpreted as leading to infinitely many gluons with small $k$ and also to power like growth of parton density with energy what violates unitarity bounds. Saturation due to nonlinear effects modifies the behaviour of distribution of gluons and in particular, since the unintegrated gluon density has a maximum, it sets the most probable momentum to be of the order of $Q_s$, what can be read off from formula (\ref{eq:poch2}). This we interpret as giving by nonlinear effects (here we study saturation model but saturation is also studied within nonlinear QCD evolution equations \cite{Kovchegov:1999yj,Kovchegov:1999ua}) emergent mass (gauge invariance still holds) to gluon which is proportional to the saturation scale and which renders the propagator to be finite.\\
We are also motivated by studies of QCD  by means of nonperturabative methods where it is argued that gluon mass consistent with gauge invariance might be dynamically generated due to nonlinearities (see \cite{Mathieu:2011mq} and references therein).
The regulated divergency of propagators due to generation of effective mass $Q_s(x)$ of virtual gluons has also impact on real gluons since they acquire mass. We can therefore write for single real gluon :
\be
M_G(x)=Q_s(x)
\ee
This observations allows us to define entropy of gluons, because the existence of temperature and existence of effective mass of gluons allows for introduction of the thermodynamical entropy (coarse grained entropy). 
Using the laws of thermodynamics and assuming small change of the volume associated with system of gluons we obtain:
\be
dE=TdS
\label{ref:termo}
\ee
and setting $dE=dM$
gives:
\be
dM=TdS
\label{eq:termo}
\ee
where the mass is understood to be the emergent mass of system of gluons carrying small longitudinal momentum fraction of the hadron and is related to mass of single gluon via:
\be
M(x)=N_G(x) M_G(x)
\ee
where $N_G(x)$ is (dimensionless) number of gluons defined as:
\be
N_G(x)\equiv\frac{dN}{dy}=\frac{1}{S_\perp}\frac{d\sigma}{dy}
\ee
where $dN/dy$ is multiplicity of produced gluons.
 Now we use eq (\ref{eq:temperatura}) what allows us to link the saturation scale to produced entropy:
\be
d\left[N_G(x)\,M_G(x)\right]=\frac{Q_s(x)}{2\pi}dS
\label{eq:entropianowa}
\ee
Thus we obtain:
\be
S=\frac{6C_F \,A_{\perp}}{\pi\alpha_s}Q_s^2(x)+S_0
\label{eq:entropia}
\ee
and equivalently
\be
S=3\pi\,\left[N_G(x)+\,N_{G\,0}\right]
\label{eq:finalfinal}
\ee
where $S_0=\frac{6C_F\,A_{\perp}}{\pi\alpha_s}\mu^2$ is a state of the lowest entropy or the minimal number of gluons.\\
Obtained entropy can be also expressed in terms of number of gluons whose distribution is given by unintegrated gluon density. The number of gluons which build up unintegrated gluon density is:
\be
n_G(x)\equiv\frac{1}{\pi}\int d^3r\,d^2k\,\Phi(x,k,r)=\frac{1}{\pi}\int d^2k\,\phi(x,k^2)
\label{eq:numberofgluons}
\ee
where $d^3r\equiv d^2b\,dl$ and $r\equiv(l,{\bf b})$, $l$ is longitudinal dimension of the proton, $b$ is already introduced impact parameter, $\Phi(x,k,r)$ is understood to 
be gluon density unintegrated in $r$, $k$, $x$.
For simplicity we assume that longitudinal space distribution of gluons factorises and is given by $\delta(l)$. This assumption is justified since the nontrivial dynamics takes place in the transversal space. Therefore we obtain:
\be
n_G(x)=\frac{C_F A_\perp}{2\pi^2\alpha_s}Q_s^2(x)
\ee
Now we can expressed entropy of produced gluons by number of gluons given by unintegrated gluon density:
\be
S=12\pi\,n_G(x)+3\pi\,N_{G\,0}
\label{eq:final2}
\ee 
This can be used perhaps to define via (\ref{eq:final2}) entropy of gluons described by unintegrated gluon density, however we postpone this task for future studies.
Although, we used GBW model to arrive at formula (\ref{eq:entropia}) our result is model independent in saturated region since GBW model is consistent with results of\cite{McLerran:1993ni,Kharzeev:2003wz,Kovchegov:1999yj,Kovchegov:1999ua,Iancu:2003ge} where gluon density in saturated region behaves like:
\be
\phi(x,k^2)\sim \frac{C_F A_\perp}{\alpha_s}\frac{k^2}{Q_s^2(x)}
\label{eq:mclarren}
\ee
The results we obtained above and in particular (\ref{eq:entropia}) are valid when $Q_s\!\!>\!\!>\mu$ where $\mu$ is an unknown constant, we choose this constant to be $\Lambda_{QCD}$. In order to obtain a formulation valid also for small temperatures one has to go along lines of the Nernst theorem, 
going along these lines, one can expect that the lowest possible entropy will be zero  what will correspond to situation where coherence of a proton \cite{Good:1960ba} is not affected by interactions.
Furthermore, we see in agreement with intuition that the entropy is proportional to number of gluons and depends linearly on the hadron size i.e. doubling the hadron size the entropy is doubled.\\
Here we would like to stress that our construction of entropy given by (\ref{eq:finalfinal}) is quite general.  The crucial element is the existence of saturation which links the temperature and the mass of system of gluons allowing for introduction of thermodynamical relations between these quantities.\\ There is one more consequence of the dependence of entropy on the saturation scale. It gives upper bound on the amount of produced entropy associated with gluons confined in saturated region of phase space what follows from our result combined with result of \cite{Albacete:2008ze,Avsar:2010ia}. In these investigations authors found that at extreme energies saturation scale eventually saturates what in combination with (\ref{eq:entropia}) gives:
\be
\Delta S_{max}=3\pi\,\left[N_{G\,max}(x)+\,N_{G\,0}\right]
\label{eq:entropia2}
\ee
Entropy however can still be produced outside of saturated region even if saturation scale will saturate itself. This we can  guess from approaches which take care for more realistic behaviour of gluon density outside of saturation region.
Namely outside of the saturation region i.e. $k>Q_s$ the gluon density from GBW model falls too fast (due to exponential behaviour it can be integrated to infinity) while BK or McLerran-Venugopalan \cite{McLerran:1993ni} model gives hard perturbative behaviour of type $Q_s^2/k^2$ which gives logarithmic divergent contribution after integrating what will give contribution to entropy unbounded by saturation of the saturation scale.    
%%%%%%%%%%%%%%%%%%%%%%%%%%%%%%%%%%%%%%%%%%%%%%%%%%%%%%%%%%%%%%%%%%%%%%%%% 
\section{Conclusions}
In this letter we have studied properties of the gluon distribution possessing the feature of saturation. We defined the saturation scale mathematically as a maximum of the gluon density (\ref{eq:poch2}). We have calculated the cross section for the inclusive gluon production and have shown that it might be approximated by saturation scale characterising gluon density of less saturated hadron. Furthermore, we introduced saturation scale related entropy of produced gluons and expressed it in terms of multiplicity of produced gluons. We have shown that this entropy behaves like number of gluons whose distribution is given by unintegrated gluon density.
We also concluded that since saturation scale ultimately saturates itself, there should exist an upper bound on entropy of gluons coming from saturated part of gluon distribution. The result we obtained is of course a contribution to the entropy coming just from emergence of saturation scale and effective mass of gluons in perturbative regime.\\ 
\section*{Acknowledgements}
I would like to thank Krzysztof Golec-Biernat and Andrzej Bialas for comments and interesting discussions. Also discussions with Wojciech Florkowski, Piotr Surowka, Eddi de Wolf are kindly acknowledged. Finally I would like to thank Yuri Kovchegov for useful correspondence.\\
This work has been supported by {\bf Fundacja na rzecz Nauki Polskiej} with grant {\bf HOMING PLUS/2010-2/6}.

\end{document}